# Sunda Arc seismicity: continuing increase of high-magnitude earthquakes since 2004

Nishtha Srivastava[1,3]*, Jonas Köhler[1,3], F. Alejandro Nava[6], Omar El Sayed[1,2], Megha Chakraborty[1,3], Jan Steinheimer[1,2], Johannes Faber[1,2], Alexander Kies[1], Kiran Kumar Thingbaijam[5], Kai Zhou[1,2], Georg Rümpker[1,3], and Horst Stoecker[1,2,4]

[1] Frankfurt Institute for Advanced Studies, 60438 Frankfurt am Main, Germany

[2] Institut für Theoretische Physik, Goethe- Universität Frankfurt, 60438 Frankfurt am Main, Germany

[3] Institute of Geosciences, Goethe- Universität Frankfurt, 60438 Frankfurt am Main, Germany

[4] GSI Helmholtzzentrum für Schwerionenforschung GmbH, 64291 Darmstadt, Germany

[5] GNS Science, Avalon, Lower Hutt 5010, New Zealand

[6] Centro de Investigación Científica y de Educación Superior de Ensenada, Ensenada, B.C., Mexico

*srivastava@fias.uni-frankfurt.de

**Abstract:**

Spatial and temporal data for earthquakes with magnitude M ≥ 6.5 can provide crucial information about the seismic history and potential for large earthquakes in a region. We analyzed ~313,500 events that occurred in the Sunda-arc region during the last 56 years, from 1964 to 2020, reported by the International Seismological Center. We report a persistent increase in the annual number of the events with $m_b$ ≥ 6.5. We tested this increase against the null hypothesis and discarded the possibility of the increase being due to random groupings. The trend given by Auto-Regressive Integrated Moving Average suggests continuing increase of such large-magnitude events in the region during the next decade. At the same time, the computed Gutenberg-Richter *b*-value shows anomalies that can be related to the occurrence of the mega 2004 Sumatra earthquake, and to possible state of high tectonic stress in the eastern parts of the region.

## Introduction

Earthquakes are a major menace among all the natural hazards, affecting many countries and resulting in significant human losses each year. A single extremely strong earthquake can take a toll of several hundred thousand lives; an example is the 2004 *Mw* 9.2 Sumatra earthquake. Large earthquakes can also trigger ecological disasters if they occur close to a dam or a nuclear power plant; the 2011 Tohoku earthquake caused a tsunami that drowned ~20,000 people (as reported by the National Police Agency of Japan) and destroyed the Fukushima Dai-ichi nuclear power plant. These catastrophic events are collectively responsible for not only over

500,000 fatalities but also for destruction of social infrastructure and heritage sites, leading to economic damages exceeding US $200 billion [1].

In spite of numerous observational studies, theoretical modeling, and laboratory simulations, precursors to the occurrence of strong earthquakes are yet to be deciphered [1]. Nonetheless, impetus on the research on the precursors is hugely significant. An improved understanding of earthquake occurrences will benefit early warning, rapid response and risk mitigation efforts, especially for developing sustainable human habitats in earthquake-prone regions.

The present study focuses on temporal patterns of earthquakes that occurred in the Sunda Arc region over the last ~60 years. The interactions of the Eurasian, Indo-Australian, Philippine and Pacific plates causes intense seismicity, volcanism and active orogeny in the region (see [2,3]). The interplate velocities along Benioff zones reach at least 10 cm/yr along Sumatra and Java giving rise to the subduction zone, the magmatic zone and the foreland basin[3]. The southwest Sumatra subduction zone is part of a long convergent belt extending from the Himalayan front southwards through Myanmar, continuing south past the Andaman and Nicobar Islands and Sumatra, south of Java and the Sunda Islands (Sumba, Timor), and then wrapping around towards north [4].

In the present analysis, we use the earthquake catalog from the International Seismological Center. We adopt a classical Machine Learning approach, for the statistical analysis and forecasting the frequency of strong events, reported yearly from 1964 to 2020, as a function of time. Since it is well known that the body-wave magnitude $m_b$ saturates for magnitudes 6 or larger (e.g. [5,6]), we test whether it is a viable tool to study changes in seismicity and in the time behavior of the Gutenberg-Richter *b*-value. We investigate to see whether, without hindsight, a clear anomaly would have been observed regionally before the 2004 $M_W$ 9.2 Sumatra earthquake. Furthermore, we study the Sumatra and the complementary Indonesia East sub-regions separately, in order to interpret the regional observations and to derive conclusions about possible current tectonic stress distribution.

## Spatial Distribution and Magnitudes of the Earthquakes

The study region is located near the equator from 13° South to 11° North and from 92° to 166° East, and will be referred to as Indonesia. The seismic events in the Sunda-Arc region have been recorded since the year 1964 onwards and are reported by the International

Seismological Center (ISC, http://www.isc.ac.uk , last accessed in January 2021). The dataset comprises latitude, longitude, origin time, focal depth and magnitude data for ~313,500 events.

The ISC-catalog has predominantly reported body-wave magnitudes in this region. Hence, we base the temporal analysis on this type of magnitude. Events reported without information on $m_b$, spatial coordinates and temporal information are excluded from the analysis.

Saturation effects associated with the body wave magnitudes ($m_b$) are certainly relevant for $m_b \geq 6$ and may lead to an underestimation of earthquake strength as measured by energy release or rupture area [5,7–9]. There are also potential limitations due to the quality of the recorded data, methods and guidelines followed in the estimation of $m_b$, apart from distributions of the reporting stations [9]. Previous researchers investigated the reliability and consistency of the ISC catalog, and concluded that reports for events with $m_b > 4.5$ are reliable [9,10].

The spatial distribution of the seismic events with reported $m_b \geq 4.5$ is shown in **Figure 1** along with the pie-charts of focal depth distribution. The distribution of the focal depth of the seismic events for each magnitude range is shown with different colors in addition to a pie-chart for better understanding of focal depth distribution in the region.

On the west, the seismicity occurs along the subduction zone, formed by the collision of the Eurasian and the Indian plates, the so-called Sunda Trench. Parallel to the Sunda Trench is the ~1900 km long Sumatran Fault, which has a long history of many damaging earthquakes[4]. On the east, active deformation takes place within a complex Suture Zone (linear belt of intense deformation) which includes several relatively small faults and subduction zones [11].

Figure 1 shows a predominance of shallow seismic events with focal depths in the range of 0-70 km. This is observed for all magnitude ranges and is likely related to the more brittle nature of the crust in shallow regions. Although the events appear more or less uniformly distributed for focal depths below ~250 km, two separate clusters with deep focal depths are observed beyond this depth. One of the two clusters represents the tectonic activity of the plates of the Java region while other deep earthquakes are due to the complex tectonics along the line of collision of the Sunda plate with the Philippine Sea plate and the Caroline plate (see [4]). Two narrow clusters of high magnitude events with $m_b > 6$ are also observed below 500 km depth.

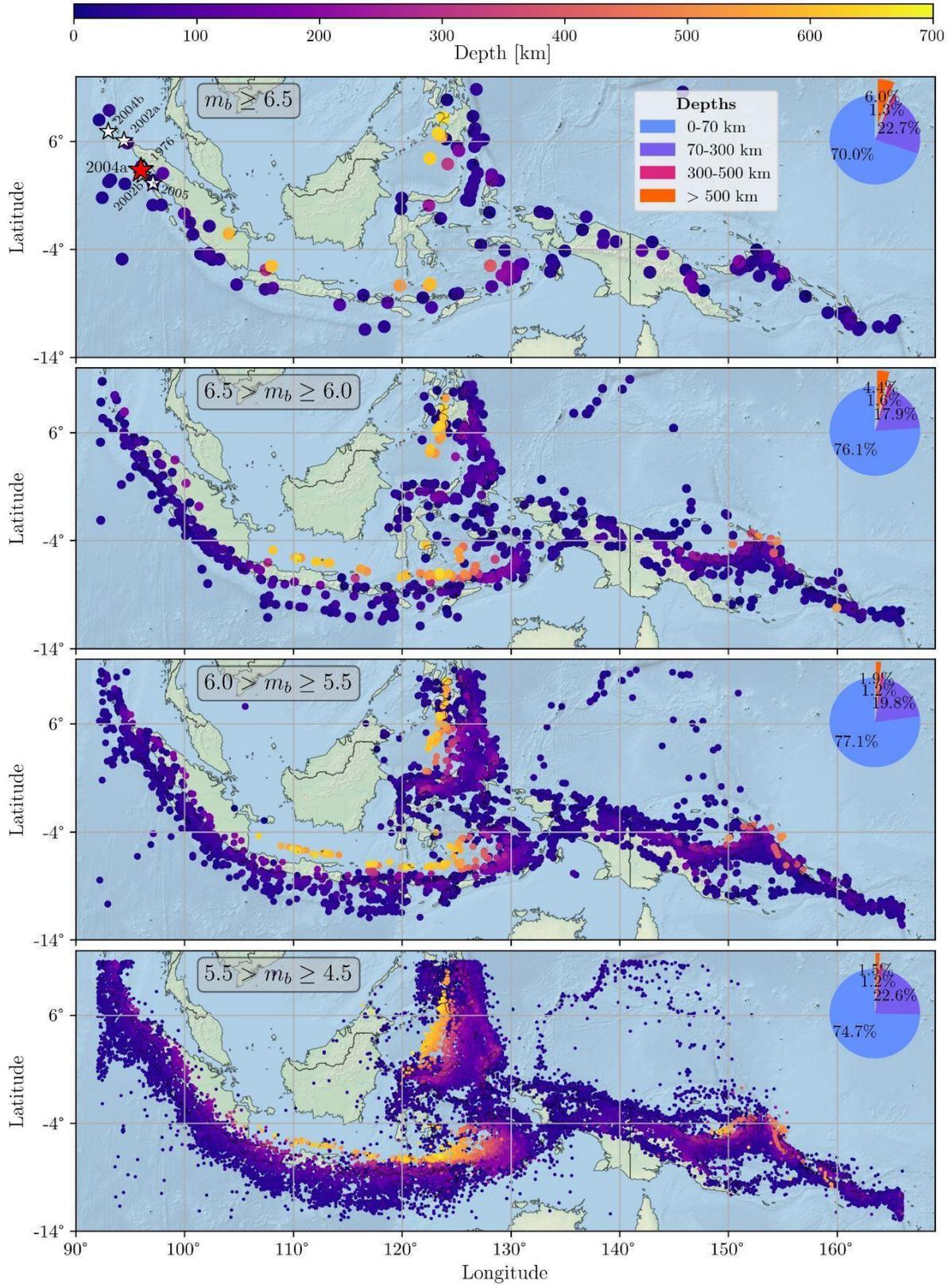

**Figure 1**: **Spatial Seismicity Variation**: Distribution of seismic events reported during 1964-2020 by the International Seismological Center for magnitudes in the ranges of $m_b \geq 6.5$, $6.5 > m_b \geq 6.0$, $6.0 > m_b \geq 5.5$ and $5.5 > m_b \geq 4.5$ (in the top panel to the bottom one). At least 70% of the reported earthquakes are shallow with focal depth in the range of 0-70 km. The epicenters of major earthquakes that occurred in the Sumatra region are shown as stars. The red star depicts the great Sumatra earthquake.

## Temporal Observation and Analysis

Due to the establishment of new seismographic stations, better global coverage and improved availability of seismic data, ISC has an improved reporting of seismic events [12]. Figure 2 shows the timelines of the steep growth in the reported numbers of events (with $m_b \geq 3.5$).

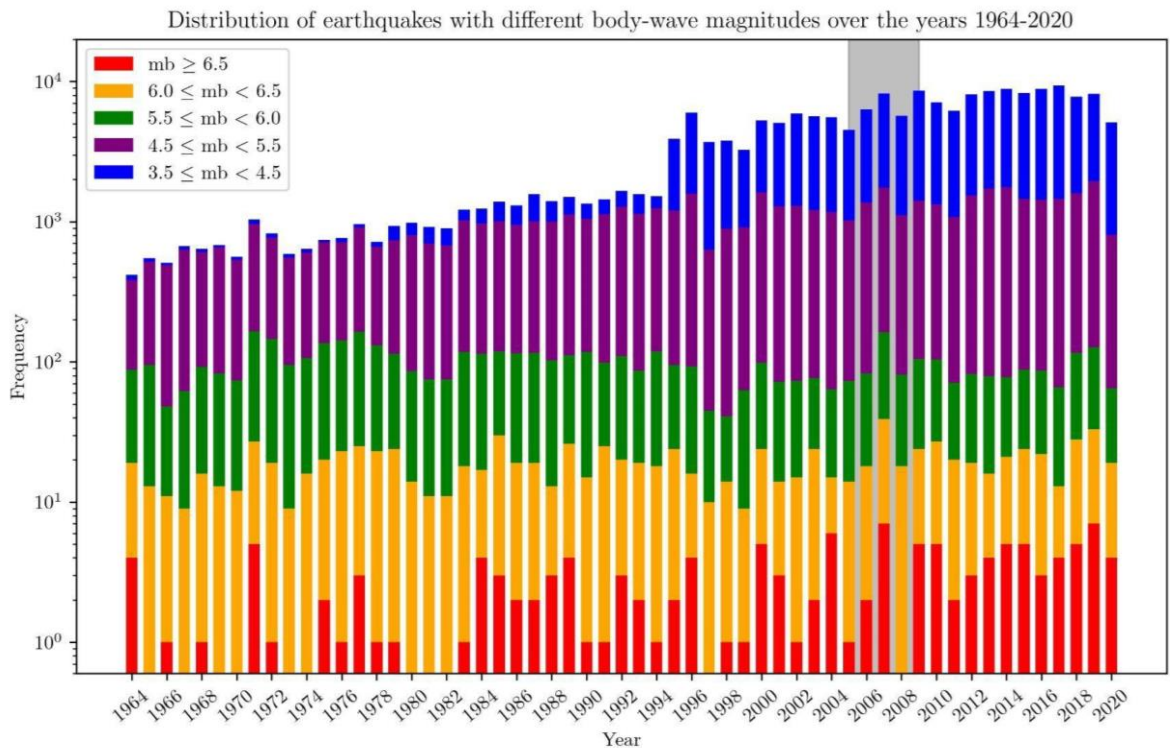

**Figure 2**: Number of earthquakes reported between 1964-2020 in the study region. The shaded gray region represents the time-duration (between 2005 and 2009) where a change in procedure to estimate body wave magnitude might have happened. Data were obtained from the International Seismological Center website (http://www.isc.ac.uk )

An increase in small magnitudes $m_b < 5.5$ can be seen clearly, as expected from an improving network of seismic stations. At the same time, there has been a steady level of seismicity for magnitudes ranging $5.5 \leq m_b < 6.5$. It suggests that the network coverage for magnitudes above 5.5 is adequate over the whole observation time. However, events with magnitudes $m_b \geq 6.5$ show intriguing temporal clustering around 1983 and increasing in frequency in recent years.

Before investigating this apparent increase, we refer to the plot of yearly maxima in Figure 3(a) to point out that the $M_W$ 9.2 mega-earthquake of 2004 and the great $M_W$ 8.2 earthquake of 2005 appear both as $m_b$ 7.0 events. It is clear that up to 2005 the body wave magnitudes saturated at 7.0. After 2008 seven events with magnitudes between 7.1 and 7.5 are reported in the catalog. Since none of these earthquakes were larger than the 2004 mega-quake, these larger $m_b$ estimates indicate a change in the procedure to estimate body wave magnitude happened between 2005 and 2009. If this change had affected events of all sizes, it would have made the latter data incompatible with that before 2006. However, Figure 2 shows no reduction in the numbers of earthquakes with magnitudes in the $6.0 \leq m_b < 6.5$ range., This observation suggests that the new evaluation only expanded the scale for earthquakes above 6.5 (also see, [13]). Hence, we assume that analyses based on all magnitudes above 6.5 grouped together will not be severely affected by the change in scale. Nevertheless, we also consider other possible effects in trend estimation and b-value analysis.

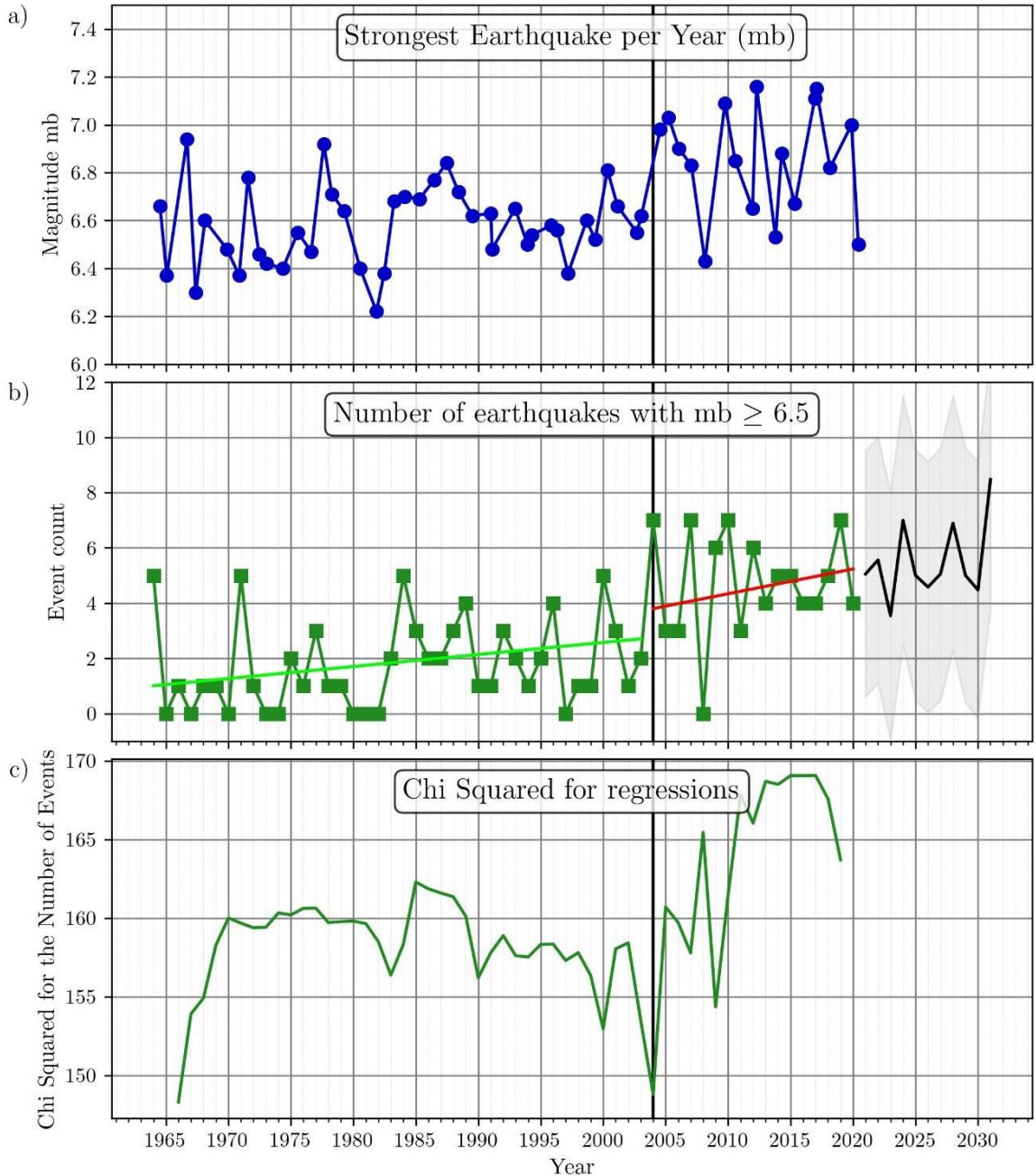

**Figure 3**: (a) The reported yearly maximum body-wave magnitude against the corresponding years. (b) As in (a), but for the number of earthquakes with body-wave magnitude $m_b \geq 6.5$. A clear increase is observed in the number of events in recent years. The gray area shows the 95% confidence band associated with the forecasting. (c) Chi-square estimates for the best fit of data ranges for the two trendlines in (b)

Another inspection into the temporal statistics of strong earthquakes is performed through the computed yearly number of events with $m_b \geq 6.5$. The number of $m_b \geq 6.5$ events seems to have doubled after ~1984, and increased substantially since 2004. For a qualitative assessment we apply the Auto Regressive Integrated Moving Average (ARIMA) model on the time series.

ARIMA is a machine learning method widely applied to capture temporal structures and in time series forecasting. This model is governed by three distinct parameters which account for seasonality, trend and noise in the datasets: *p* (the autoregressive part), *d* (the integrated part), and *q* (the moving average part), respectively. In short, *p* incorporates the effects of previous values, *d* embodies the amount of difference (the number of times the difference with the past values is considered), and *q* gives the error in the model as a linear combination of the error observed for previous values [14].

The first step of the computation is to estimate optimal values of the *p*, *d* and *q*. Here, we implement a grid search to explore various combinations of the parameters. The optimum set of parameters is decided by considering the best performance indicated by Akaike Information Criterion (AIC). AIC provides a measure that indicates how well a model fits the data while respecting the overall complexity of the model [15]. A model that fits the data using a lot of features has a larger AIC score than a model which uses fewer features to achieve the same fit quality. The best set of parameters, *i.e.* those which produced the best fitting model with lowest AIC score, was selected by grid search.

The overall increasing trend-lines seen in Figure 3 may be explained by poor reporting in the past and improved regional coverage in later years. However, the growing numbers of strong events $m_b \geq 6.5$ from the year 2004 onwards can be clearly observed. One could interpret that the seismic activity in the region has significantly increased since 2004.

Figure 3c shows the linear least squares plot for fitting two trendlines to the number of those larger earthquakes. The x-axis determines the year, in which the first trendline ends and the second one starts. Apart from low chi squared values for the border case, where one trendline connects only the first two years, the best fit splits the dataset at 2004, with the two resulting lines as shown plotted in Figure 3b.

## Null Hypothesis Test

To determine the level of significance for the observed clustering of the larger magnitude events, we consider the null hypothesis that seismic events with magnitudes $m_b \geq$ 6.5 occurred randomly in time. If the occurrence of large magnitudes were random in time, the number of events *n* in any given year should have a Poisson distribution with probabilities given by,

$$\Pr(n) = \frac{\lambda^n e^{-\lambda}}{n!} \circ p_n,$$ (1)

where $\lambda$ is the rate of events/year (e.g. [16]).

Figure 4 compares the observed number of events in each year to the corresponding expected number of events from a Poissonian distribution with $\lambda = 2.63158$ corresponding to 150 events in 57 years.

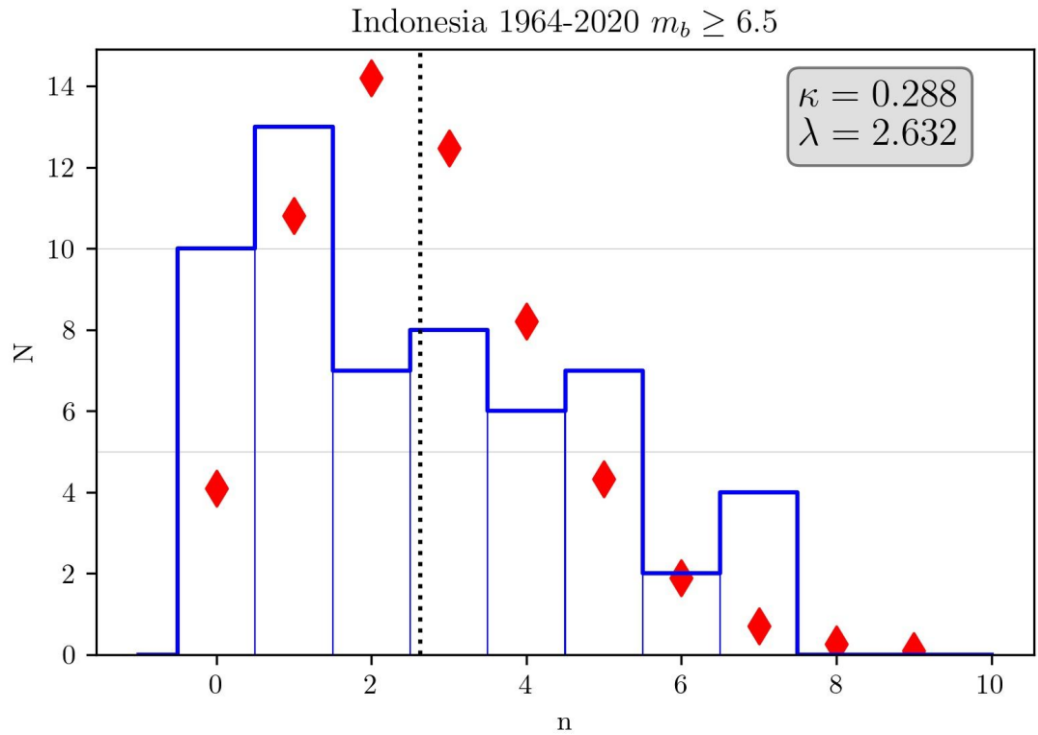

**Figure 4**: Histogram of observed number of events with $m_b \geq 6.5$ for each year from 1964 to 2020 (blue line) and corresponding Poissonian expected number of events (red diamonds). The dotted vertical line indicates the value of $\lambda$; $\kappa$ is the Kullback-Leibler divergence.

To quantify how much the observed distribution differs from the Poissonian one, we used the Kullback-Leibler divergence (K-L) as follows:

$$k = \sum_{n=0}^{n_x} p_n \log_2\left(\frac{p_n}{\rho_n}\right)$$ (2)

($^{17}$), where $\eta_x$ is the maximum number of observed events in a year, and $p_n$ is the observed probability of having $n$ events in any one year, as given below,

$$p_n = \frac{N_n}{\sum_{j=0}^{n_x} N_j}. \tag{3}$$

Note that the summation in (2) has been restricted to the range where $p_n$ is not zero, since zero values do not contribute to the sum. The K-L divergence is zero when $P = \Pi$, but it does not have a fixed upper limit (see $^{18}$). To determine how significant the observed $\kappa = 0.28756$ is, we conducted a Monte Carlo simulation of 50,000 realizations of 57 years samples of Poisson distributed numbers and obtained the $\kappa$ histogram as shown in Figure 5. The cumulative histogram shows that Pr ($\kappa \geq 0.2875$) = 0.00510, confirming that the null hypothesis can be rejected with ~0.995 confidence.

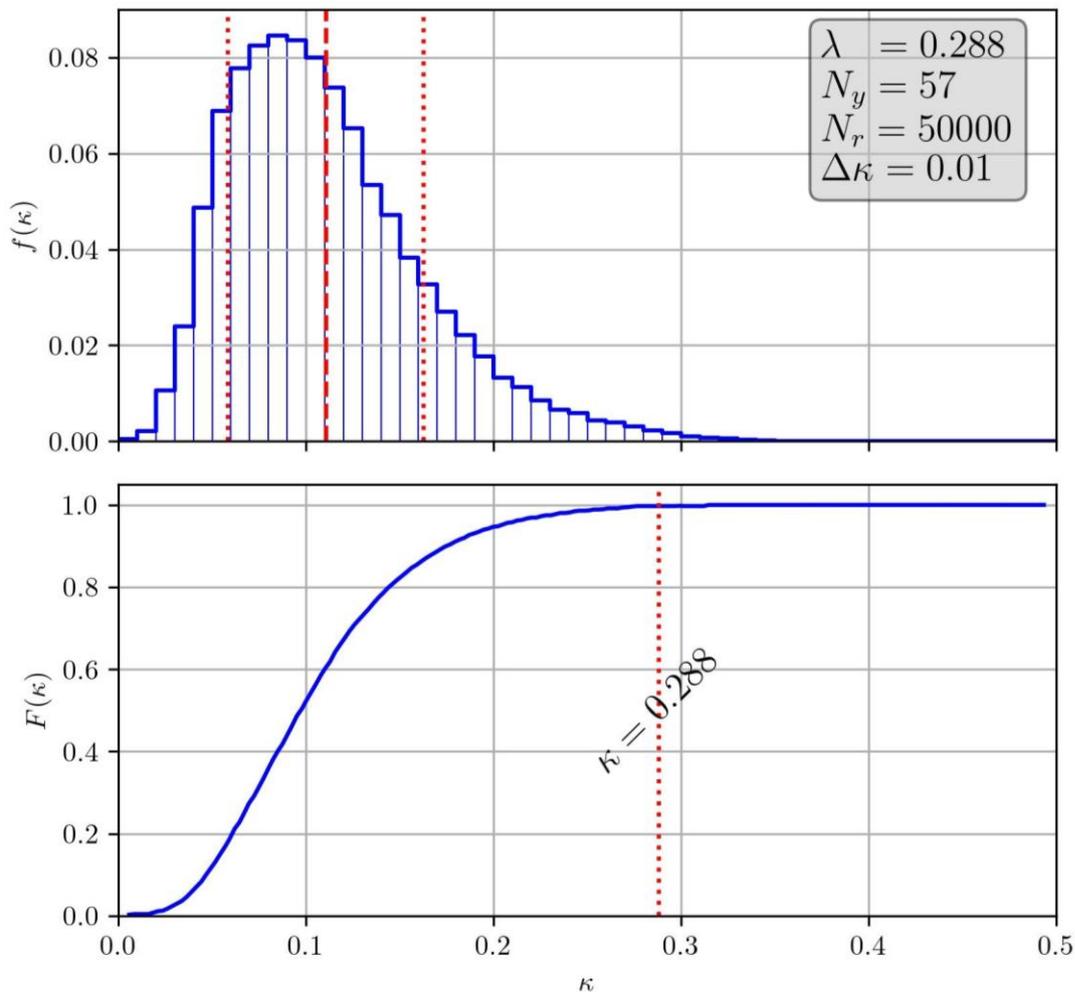

**Figure 5**: Histogram of values from the Monte Carlo simulation (top) the dashed vertical line indicates the mean of the $N_r = 50000$ realizations (0.11082), and the dotted vertical

lines indicate plus/minus one standard deviation (0.05219) from the mean. Cumulative κ distribution (bottom), the dotted line indicates the observed κ, which is located 3.39 standard deviations away from the mean.

To examine whether this result was dependent on including events after the scale change, we carried out the same exercise using a subset of the data covering the period from 1964 to 2008. This subset of the data has 90 events in 45 years that corresponds to λ = 2.0 events/year and yielded κ = 0.26284. The Monte Carlo estimate gives Pr (κ ≥ 0.26248) = 0.02247, and hence, the null hypothesis can still be rejected, but with 0.978 confidence.

## b-value Analysis

After discarding the possibility of the increasing frequency of large magnitude earthquakes being random, we proceed to investigate the variations in the frequency magnitude distributions. A widely used measure of the large magnitudes to small magnitudes ratio is the b-value[19,20]. This parameter has been identified as an observable precursory to large earthquakes (e.g. [21–23]). It is widely accepted to be inversely related to the state of tectonic stress in seismogenic areas (e.g., [24–27]) which explains observed low-value anomalies preceding many large earthquakes.

To estimate b-values, we use the Nava et al. [28] method. This method determines the most likely b-value bx that results in an observed Aki-Utsu b-value [29,30] for some given magnitude range and number of data in a G-R histogram. The present analysis is based on mb magnitudes, and hence, the estimated b-values may not agree with those based on MW magnitudes.

A straightforward *b* versus time analysis using 10 year-long time windows (in Figure 6a) shows *b* going from large to small *b*-values with two strong jumps sometime around 1980 and 1990, reaches a minimum for 1991 to 2000, increases a little (within the uncertainty bounds), and decreases again for 2011 to 2020. These results conforms to increasing tectonics in the region.

Since the 2004 mega-earthquake could impact the estimated b-values, a logical way to configure the time windows is with reference to this earthquake, as shown in Figure 6b. This configuration tells a slightly different story: b-values decrease gradually from 1964 until the occurrence of the two large earthquakes in late 2004 and early 2005, and then continue decreasing more slowly until before 2021. The trend in the b-values remains overall the same, supporting the likelihood of growing tectonic stresses in the region. Does it mean that another, even larger, earthquake should be expected soon?

Since the region under study is quite complicated, considering it as a whole may lead to erroneous conclusions. Hence, we decided to focus on two sub-regions separately. The first one, which we refer to as the Sumatra region, corresponds to the ruptures of the 2004 and 2005 earthquakes, and comprises the western part of the study region from 92°E to about 100°E. The second one comprises the rest of the region and is referred to as Indonesia E.

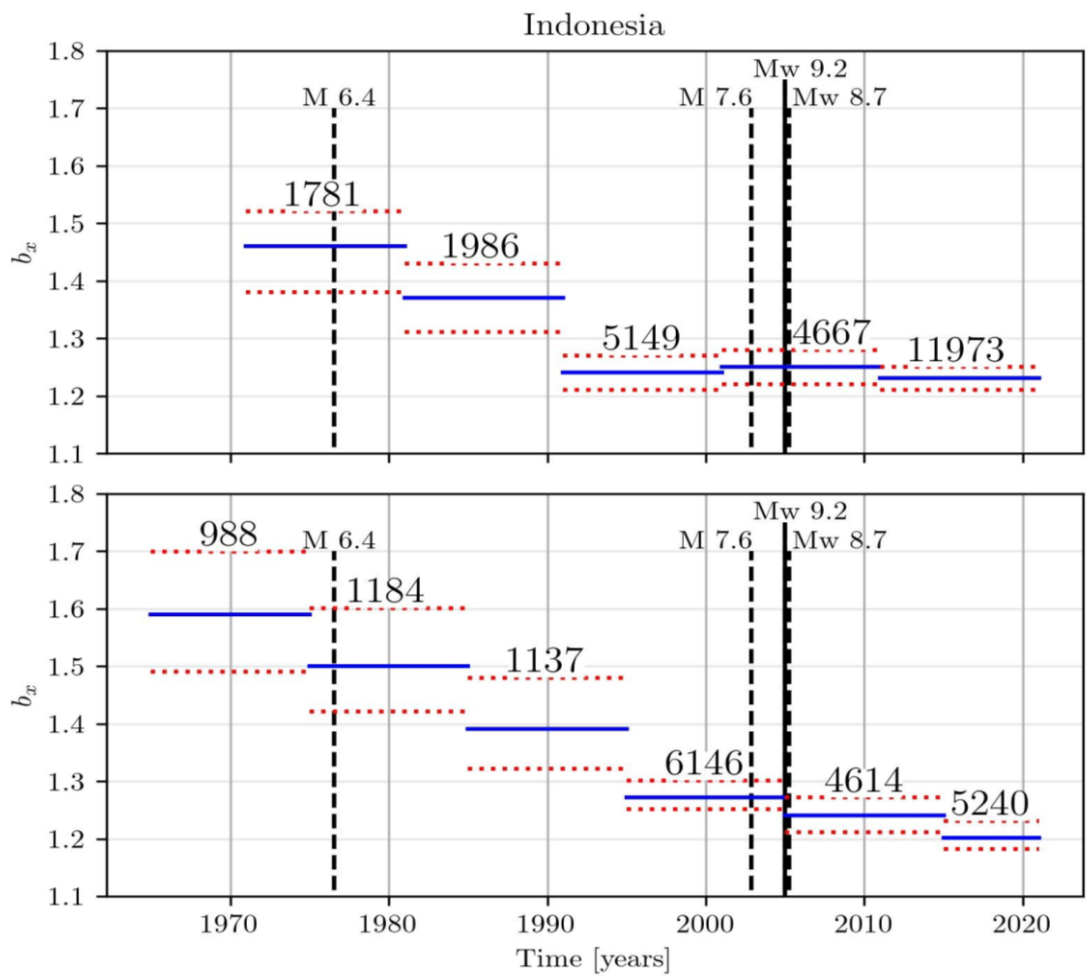

**Figure 6**: Most likely source $b$-values, $b_x$, vs. Time, for the Sunda Arc region and different 10 year long windows. The $b_x$ values are indicated by thick horizontal lines extending over the corresponding sampling time window; the dashed lines above and below each $b_x$ value indicate the 90% confidence range, and the numbers above each line are the number of earthquakes in the linear G-R range for the time window. The vertical lines indicate the occurrence times of the 2004, 2005 and other two large earthquakes.

The results for the analysis of the separate sub-regions are shown in Figure 7; the smallest length of the time windows is five years, but before 1996 for Sumatra and before 1984 for Indonesia E, it was necessary to use longer time windows to have enough data for reliable *b*-value estimations [31,32], and 1 year overlaps were used for continuity.

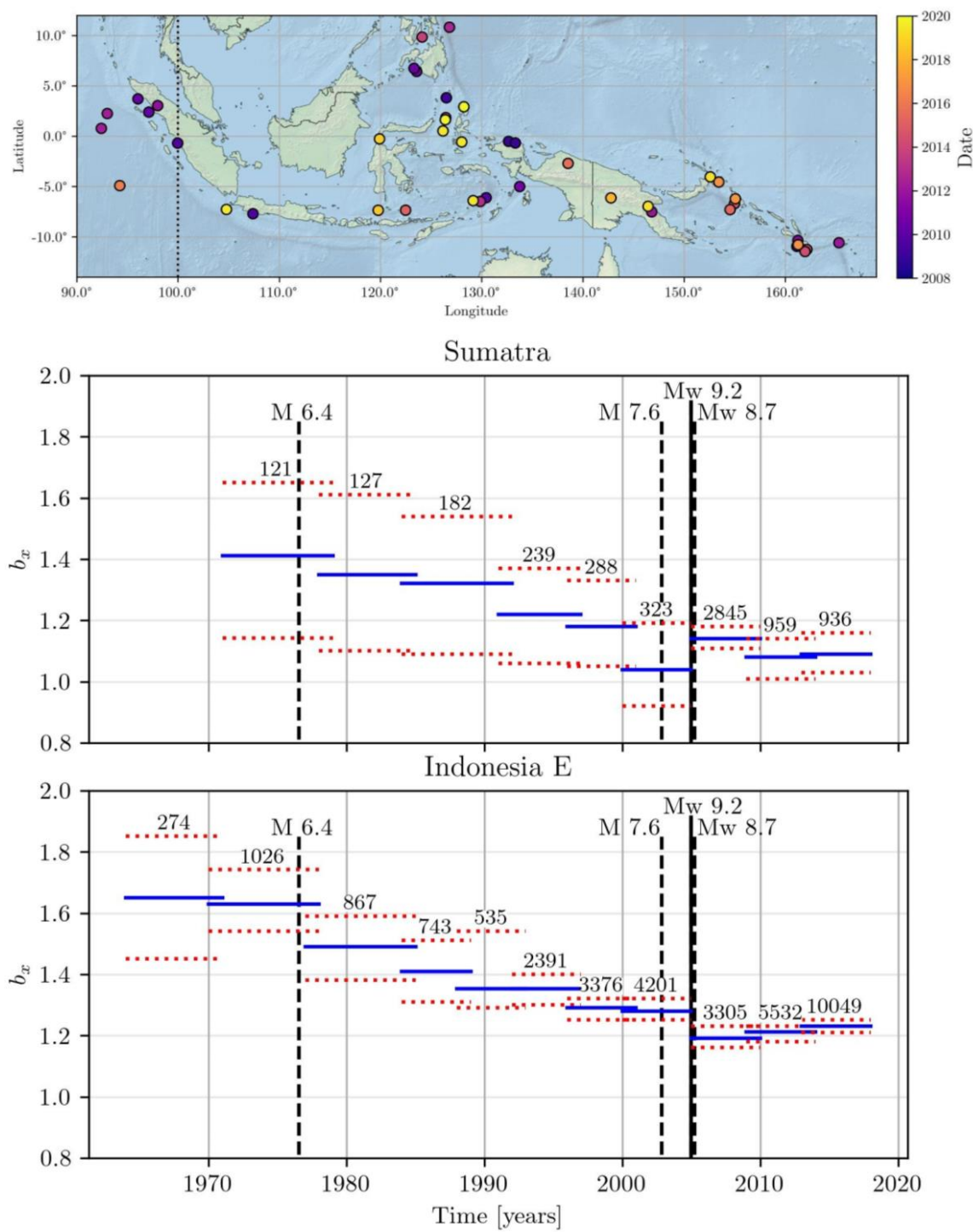

**Figure 7**: Top figure shows the epicentral locations of major earthquakes. Most likely source $b$-values, $b_x$, vs. time, for the Sumatra (middle figure) and Indonesia E (Below figure) sub-regions for the same 10 year-long time windows. The $b_x$ values are indicated by thick horizontal lines extending over the corresponding sampling time window; the dashed lines above and below each $b_x$, value indicate the 90% confidence range, and the numbers above each line are the number of earthquakes in the linear G-R range for the time window. The vertical lines indicate the occurrence times of the 1976, 2004 and 2005 earthquakes.

# Discussion

Before the 2004 earthquake, $b$-values in the Sumatra sub-region (Figure 7) are consistently lower than those in Indonesia E and in the whole region (Figure 6) which indicates that the higher stresses were at that time found in the Sumatra sub-region. For this region, $b_x$ decreases from 1.41 for 1971 to 1978 to a minimum of 1.04 during the five years before the 2004 earthquake, which is within the precursory time lapse observed for other earthquakes (e.g. [33]), then increases as would be expected for a region where stress has been largely liberated [33–35], to 1.14, and surprisingly decreases to 1.08 for 2009 to 2013 and then increases slightly to 1.09; although the 90% confidence limits of these last 3 values do overlap a bit, we propose that the largest decrement around 2009 was due to the previously mentioned change in the magnitude scale, so that this decrement does not indicate that stresses continue to increase in the Sumatra sub-region.

The three decades decrease in $b$-value for the Sumatra sub-region before the 2004 earthquake, shown in Figure 7, if monitored through some kind of foresight, could have been a valuable indicator of the possible forthcoming of a large earthquake. This result agrees with other studies of $b$-value anomalies before this earthquake. Nuannin et al.[6] present $b$-value determinations for this sub-region over a time corresponding to our 2000 to 2004 window and found slightly smaller values, around 0.7 to 1.0, using $M_S$ and much larger values, around 1.2 to 1.6, using $M_W$, their determinations were done through direct fitting of the spectra but they used only 50 events for each of their determinations, so that considering the expected uncertainties, their $M_S$ $b$-values agree with our results.

Dasgupta et al.[36] estimated spatial and temporal $b$-values before the 2004 earthquake, using $m_b$ (ISC for 1976-2001 and NEIC for 2002-2004), spatio-temporal windows were 2°×2°×3 years containing "a minimum of 10 events", but what were the actual numbers of events used for each determination is not known; the temporal variation in the epicentral area was obtained from the average of the values in the two blocks closer to the epicenter. They found a rise in *b*-value from about 0.83 to a maximum of about 1.16 between 1993 and 1998, and a decrease to about 0.55 between 1998 and the time of the 2004 earthquake.

Roy et al. [37] mapped *b*-value and fractal dimensions in the Andaman-Sumatra subduction zone for events occurring from 1964 to 2007 and found zones of low *b*-value, particularly at depth, in sections crossing the trench at longitudes ~102°E and 111°E, which could indicate ongoing high stresses at Indonesia E; unfortunately they do not state sample sizes.

Nanjo et al. [21] found pronounced decade-scale decreases in *b*-value for both the 2011 Tohoku and 2004 Sumatra earthquakes. For Sumatra, using the ANSS catalog, they found values decreasing from ~1.4 some 40 years before the 2004 earthquake to ~1.03 some 5 years before it; they do not state which magnitude or how many events were used, but their results are consistent with ours.

Now, let us look at the *b*-values in the complementary Indonesia E region; these values decrease consistently from, $b_x = 1.65$ during 1964 to 1974, to $b_x = 1.28$ for the 1995 to 2004 time window, they do not increase after the 2004 earthquake but show an all time low of 1.19 for the time window right after it, then increase slightly to 1.21 and 1.23 during the following two time windows. It may be speculated that the decrease after the 2004 earthquake was due to stress concentrations caused by it, concentrations that afterwards redistributed obscuring the effects of the change in magnitude scale.

It should be pointed out that *b*-values in Indonesia E are still low, not far from the critical level found for Sumatra, which indicates high stresses in this sub-region, as evidenced by the high occurrence of large magnitudes shown in Figure 2. We propose that it is important to monitor this region, both as a whole and focusing in particular sub-sub-regions, to identify possible stress concentrations as places with high seismic hazard.

The 26 December 2004 Sumatra-Andaman earthquake is considered to be one of the largest seismic events followed by a devastating tsunami in recorded history [38]. This event occurred at a focal depth of approximately 30 kms and ruptured ~1300 km of the Indo-Australian subduction zone curved plate boundary from northwestern Sumatra to Andaman Island. One of the strongest aftershocks of this event, which triggered on 28 March 2005, ruptured the adjacent 300 kms [38]. This highly active tectonic regime also appears to be host to some great historical earthquakes which are reported to be triggered to the southeast of the

2004 rupture zone in the years: 1797, 1833 and 1861. Additionally, two great earthquakes were also reported in the year 1881 and 1941 thus establishing the threat of great earthquakes for the region in the future as well [38–42]. Lay et al [38] observed that 40 years prior to the 2004 event little seismicity occurred within 100 km of the trench between the epicenters of the large events triggered in the year 2004 and 1881 thus suggesting a long-term strain accumulation in the eventual rupture zone. The average increase of the number of events with mb ≥ 6.5 is also observed in the trendline of Figure 3(b).

The noticeable increase of large events from 2000 prior to the 26 December 2004 earthquake could be an indicator of accelerating moment release (AMR) pattern in seismicity which has been an intriguing topic for various theoretical and observational studies [43–55].

A study by Nishenko and McCann [56] claims that the inner structure variations on the inner wall of trenches appear to reflect changes in both the lengths of rupture zones and in the source areas of tsunamis that are associated with large shallow earthquakes. Upper slope basins, deep sea terraces and other topographic features may act as an indicator of the tectonic rate and seismic-tsunami risk along convergent plate margins [39]. With this thought in mind, one potential reason for the increase of earthquakes with mb ≥ 6.5 after the 2004 earthquake could be an opening of a comparatively bigger lock between the subducting and overriding plate. This lock along the fault boundary could be the cause of huge stress accumulation later resulting in a great earthquake and tsunami.

Owing to the underlying complex tectonic structure of this seismogenic regime, we envisage a further extension of this study by segregating the spatial and temporal clusters of the seismicity to understand the reported phenomenon in a follow up study.

In this study we report an increase in the average number of strong earthquakes per year in the Sunda Arc as reported by ISC. The analysis shows that during the next decade more and stronger earthquakes may occur along the Indonesian segments of the Pacific Ring of Fire. The observed trend of seismic activity on the rise is worrisome. It would be imperative that Indonesia must prepare: earthquake-resistant buildings, improved warning systems and practical evacuation protocols must be prioritized. Machine Learning and Deep Learning based methods may become the key to fostering these efforts.

**Methods**

**b value estimation**

The Gutenberg-Richter (GR) relation [19] states that the number of earthquakes with magnitudes greater than or equal to a given magnitude $M$ is distributed as

$$\log_{10} N(M) = a_1 - b(M - M_1); \quad M \geq M_1,$$

where $M_1$ is the completeness magnitude below which the Gutenberg-Richter (GR) histogram is not linear because of lack of coverage. The GR relation is equivalent to stating that magnitudes are distributed exponentially with mean $m$, which is related to $b$ as

$$b = \frac{\log_{10} e}{m - M_1}$$

The Aki-Utsu maximum likelihood estimate of $b$ for magnitudes rounded to $\Delta M$ (usually $\Delta M = 0.1$) is

$$b_m = \frac{\log_{10} e}{\overline{M} - (M_1 - \Delta M / 2)},$$

where $\overline{M}$ is the mean observed magnitude [29,30]. Of course $b_m$ will be a good estimate of $b$ only if $\overline{M}$ is a good estimate of $m$, but many times data are too few for a good estimate, especially considering that large magnitudes, those above a magnitude $M_2$ above which the GR histogram is no longer linear, will be over- or under-sampled, particularly for short catalogs.

The idea of the most-likely source $b$ estimate is to use Monte Carlo methods to determine which source or "true" $b$ value, $b_x$, is the one that is most likely to result in the observed $b_m$, given $M_1$, $M_2$, and the observed number of earthquakes within this range. For short samples and, consequently, short linear ranges, $b_m$ is usually overestimated, while for large sufficient samples $b_x$ usually agrees with $b_m$. For a full discussion of the method see Nava et al. [28].

**Acknowledgement**

Nishtha Srivastava appreciates the "KI-Nachwuchswissenschaftlerinnen" - grant SAI 01IS20059 by the Bundesministerium für Bildung und Forschung - BMBF. Calculations were performed at the Frankfurt Institute for Advanced Studies' GPU cluster, funded by BMBF for the project Seismologie und Artifizielle Intelligenz (SAI). Kai Zhou and Jan Steinheimer are supported by the BMBF through ErUM-Data funding and the Samson AG AI grant. Kai Zhou also thanks the GPU grant provided by the NVIDIA Corporation. Horst Stöcker acknowledges the Judah M. Eisenberg Professur Laureatus - Chair of the Walter Greiner Gesellschaft and of the Fachbereich Physik at Goethe Universität Frankfurt. Authors also appreciate the feedback received from Dr Madhusree Mukerjee.